\documentclass[twocolumn,prl,aps,superscriptaddress,longbibliography]{revtex4-2}

\usepackage[utf8]{inputenc}
\usepackage{amsmath,amssymb}
\usepackage{graphicx}
\usepackage{floatrow}
\usepackage[T1]{fontenc}
\usepackage{physics}
\usepackage{tikz,bm}
\usepackage[caption=false]{subfig}
\usepackage{epigraph}
\usepackage{appendix}
\usepackage{hyperref,cleveref}
\usepackage{qcircuit}

\newcommand{\be}{\begin{equation}} 
\newcommand{\ee}{\end{equation}}
\newcommand{\innpro}[2]{\left\langle#2\vert#1\right\rangle}

\begin{document}
\title{Towards fault-tolerant quantum computation with universal continuous-variable gates
}
\author{Sheron Blair}\email{sblair13@qub.ac.uk}
\affiliation{Centre for Quantum Materials and Technologies, School of Mathematics and Physics, Queen's University Belfast, University Road, Belfast BT7 1NN, United Kingdom}
\affiliation{Irish Centre for High-End Computing (ICHEC), University of Galway, Tower Building,
Grand Canal Quay, Dublin 2
D02 HP83, Ireland}
\author{Francesco Arzani}
\affiliation{DIENS, École Normale Supérieure, PSL University, CNRS, INRIA, 45 rue d’Ulm, Paris 75005, France}
\author{Giulia Ferrini}
\affiliation{Department of Microtechnology and Nanoscience (MC2), Chalmers University of Technology, SE-412 96 G\"{o}teborg, Sweden}
\author{Alessandro Ferraro}
\affiliation{Dipartimento di Fisica ``Aldo Pontremoli'',
Università degli Studi di Milano, I-20133 Milano, Italy}
\date{\today}

\begin{abstract}
Continuous-variable (CV) systems have shown remarkable potential for quantum computation, particularly excelling in scalability and error correction through bosonic encoding. Within this framework, the foundational notion of computational universality was introduced in [Phys. Rev. Lett. \textbf{82}, 1784 (1999)], and has proven especially successful since it allows for the identification of finite sets of universal CV gates independent of the encoding scheme. However, achieving the critical objective of fault-tolerant computation requires some form of encoding, and to date there has been no proof that these universal CV gates can lead to encoded fault tolerance. We present compelling evidence in this direction by utilizing the Gottesman-Kitaev-Preskill (GKP) encoding. Specifically, we numerically optimize the generation of GKP states from vacua using circuits comprised solely of universal CV gates. We demonstrate that these states can be attained with sufficient quality to exhibit error probabilities lower than the threshold needed to achieve a fault-tolerant memory via concatenated GKP-stabilizer codes.
\end{abstract}

\maketitle
 
In 1999, Lloyd and Braunstein (LB) introduced an approach to quantum computing based on bosonic infinite-dimensional systems \cite{BraunLloyd}, as opposed to the more conventional two-dimensional qubit systems \cite{Nielsen&Chuang}. 
This approach exploits observables with continuous spectra as information carriers \cite{Braunstein:05,  serafini2017quantum}, giving rise to the concept of quantum computation over continuous variables (CVs) \cite{Weedbrook:12, pfister2019continuous, fukui_building_2022-1}.
Over the years, this paradigm has ignited an active research field, yielding groundbreaking achievements at the forefront of quantum computing. These include the generation of entangled states comprising billions of individually addressable temporal modes \cite{yoshikawa2016invited, aghaee2025scaling}, the implementation of quantum codes attaining error correction beyond the break-even point \cite{ofek2016extending, sivak2023real, ni2023beating, brock2025quantum}, and leading-edge experiments pushing towards quantum computational advantage \cite{zhong2021phase, madsen2022quantum, Deng2023}. The abundance in nature of physical systems described by quantum CVs is also noteworthy, as evidenced by the diverse experimental platforms that have enabled these advances, including optics \cite{chen2014experimental, roslund2014wavelength, bourassa2021blueprint, asavanant_generation_2019, larsen_deterministic_2019, wang2024chip}, trapped ions \cite{Fluhmann,de2022error}, and superconducting systems \cite{KonnoGKPgeneration, jolin2023multipartite, kudra2022robust}.

Central to this CV framework is the following notion of computational universality, which we refer to as LB universality: a physical platform is deemed computationally universal if it can host a finite set of parametrized gates that allow for arbitrarily accurate approximation of any dynamics governed by Hamiltonians expressible as polynomials in the field variables, through appropriate gate sequences \cite{BraunLloyd}.
Several studies have sought to identify platforms capable of hosting an LB universal gate set \cite{Weedbrook:12, pfister2019continuous, fukui_building_2022-1, park2018deterministic, marek2018general, hillmann2020universal, houhou2022unconditional, mcconnell2022multi, park2024efficient, budinger2024all}, culminating recently with the first experimental demonstration using superconducting circuits \cite{eriksson2024universal}.

A notion of computational universality is of limited practical value if it is not equipped with a robust strategy for fault tolerance \cite{shor1996fault}, which is indispensable for achieving large-scale quantum computing \cite{gottesman2010introduction}. It is generally believed that the only viable way to achieve error tolerance in a CV system is to introduce some form of discretization (or encoding). This process parallels conventional classical computing, where information is digitized despite the underlying continuous nature of physical classical variables \cite{harris2015digital}. 
As a consequence, the concepts of \textit{encoded} universality and fault tolerance become central. In the context of CV quantum computing, this has led to the introduction of bosonic codes \cite{terhal_towards_2020, grimsmo_quantum_2021, joshi2021quantum, cai2021bosonic, albert2022bosonic, brady2024advances}. Theoretical work has primarily focused on designing specific instances of bosonic codes \cite{chuang_bosonic_1997,
cochrane1999macroscopically,  gkp, ralph_quantum_2003,
mirrahimi2014dynamically, michael2016new, grimsmo_quantum_2020}. While these efforts have been successful in their own right, they have largely bypassed the universality notion proposed by Lloyd and Braunstein, which is independent of any encoding. In particular, it seems reasonable to expect that a meaningful notion of universality would allow for universal encoded computation regardless of the specific encoding. However, to date, there has been no proof that an LB universal platform can indeed support encoded universality and fault tolerance. In this work, we address this gap, demonstrating that encoded universality can be achieved using LB universal CV gates and providing evidence that fault tolerance may also be attained.

\textit{LB universal model---}
The LB paradigm entails computing over individually addressable bosonic modes. Denoting with $\hat{a}$ and $\hat{a}^\dag$ the annihilation and creation operators of a mode, the position and momentum operators are given by $
    \hat{q}=\left(\hat{a}+\hat{a}^\dag\right)/\sqrt{2}\,$ and $
    \hat{p}=\left(\hat{a}-\hat{a}^\dag\right)/\sqrt{2}i\,$ respectively. Following Ref.~\cite{BraunLloyd}, we select a set of CV gates made up of displacements in position $X(c)$, squeezing $S(r)$, rotations $R(\theta)$, and Kerr gates $K_{\textrm{LB}}(\kappa)$, defined as
\begin{align}
\label{Xgate}
X(c)&=\exp\left[-ic\hat{p}\right]\;,  \\
\label{Sgate}
S(r)&=\exp\left[i\frac{r}{2}(\hat{q}\hat{p}+\hat{p}\hat{q})\right]\;, \\
\label{Rgate}
R(\theta) &= \exp\left[i\frac\theta2\left(\hat{q}^2+\hat{p}^2\right)\right]\;, \\
K_{\textrm{LB}}(\kappa) &=\exp\left[i\kappa\left(\hat{q}^2+\hat{p}^2\right)^2\right]\;,
\label{Kgate}
\end{align}
with $\{c, r, \theta, \kappa\}\in\mathbb{R}$.

It is known that this set of gates is LB universal for single-mode computation since unitary operators generated by polynomials of arbitrary order in $\hat{q}$ and $\hat{p}$ can be arbitrarily well approximated by sequences of the above gates \cite{BraunLloyd}. The addition of any nontrivial two-mode gate acting between two generic modes $j$ and $k$, such as $
CZ=\exp[i\hat{q}_j \hat{q}_k]$, allows for full multi-mode LB universality. For the LB model to enable computation, it is essential to specify not only a gate set but also a fiducial initial state and a set of CV measurements. Since the CV gate set described is mostly composed of Gaussian operators \cite{Ferraro05, adesso2014continuous} (with the exception of the Kerr gate), it is natural to use Gaussian inputs and measurements to avoid introducing further computational resources \cite{albarelli2018resource, takagi2018convex}. For simplicity, we will consider vacuum states as input and position-eigenstate projectors as measurements (homodyne measurements). We also include classical feed-forward in our definition of the LB model ---\textit{i.e.}, the ability to use classical communication to feed measurement results to subsequent gates.

As said, the versatile LB approach is agnostic with respect to the encoding, in the sense that logical gates can be constructed for any embedding of qubits into oscillators. Therefore, to demonstrate that the above defined model enables, in turn, \textit{encoded} universality and fault-tolerance, it is sufficient to prove this for a specific encoding. For convenience, we adopt the one proposed by Gottesman, Kitaev, and Preskill (GKP) \cite{gkp}. We will now present a strategy for achieving GKP-encoded universality using the LB universal model defined above.

\textit{LB universality implies encoded universal  computation---} Let us first recall that the approximate basis square-GKP states $\ket{\Tilde{\mu}}$, for $\mu\in\{0,1\}$, are defined as \footnote{Note that other definitions are also possible, but the most common in the literature have all been shown to be equivalent \cite{equivApproxGKP}.}
\begin{multline}
    \ket{\Tilde{\mu}}=N \sum_{n=-\infty}^{+\infty} \exp[-\frac{1}{2}\Delta^2([2n+\mu]\sqrt{\pi}\,)^2]\\ \cdot X\left([2n+\mu]\sqrt{\pi}\,\right)\ket{\Psi(0)}\,,
    \label{eq:approxgkp}
\end{multline}
where $N$ is a normalisation constant and $\Delta^{-1}$ is the width of a Gaussian envelope. Here $\ket{\Psi(x)}$ is the approximate position eigenstate with eigenvalue $x$,
\begin{multline}
    \ket{\Psi(x)}=\int_{-\infty}^{+\infty} dq \,\frac{1}{\left(\pi\Delta^2\right)^{1/4}}\,\exp\left[-\frac{1}{2}\frac{(q-x)^2}{\Delta^2}\right]\ket{q}_q\,,
\end{multline}
which is a squeezed state of width $\Delta$, with squeezing in decibels given by $s_{\text{dB}}=-10\log_{10}\Delta^2$. In the limit of infinite squeezing one obtains the so-called ideal GKP states, hereafter indicated as $\ket{\bar{\mu}}$.

The first step is to show that the CV gates in Eqs.~(\ref{Xgate}-\ref{Kgate}) can generate approximate GKP states from vacuum. This is not obvious, as there exists no analytical proof confirming this capability or specifying the quality of the approximation that can be achieved. In general, it is unknown whether an LB universal gate set can generate a dense family of states in the full Hilbert space of the system \cite{WuTarnLi2006}. One of the authors recently participated in a work showing that polynomially generated Hamiltonians can indeed generate any physical state from vacuum, with a rigorous approximation bound~\cite{ABC}. However, the constructive proof uses a universal set which is very different from what prescribed in Ref.~\cite{BraunLloyd} and whose size depends on the target accuracy. The possibility to extend those results to a \emph{finite} set of universal Hamiltonians, the size of which does not depend on the approximation error, remains an open question. As a matter of fact, while various proposals have been put forward for generating GKP states, none of these make use of an LB universal model. Some methods are probabilistic and conditioned on specific measurements \cite{gkp, Pirandola04, PirandolaPond,  PirandolaLith, vasconcelos,  Eaton_2019, su2019conversion, Tzitrin2020, eaton2021mbgridstates, Fukui, zheng2023gaussian, brenner2024complexitygottesmankitaevpreskillstates, crescimanna2025adaptivenongaussianquantumstate}, while others involve auxiliary qubits \cite{TravMilburn,hastrup_npj, Motes2017, TerhalWeigand}, or higher-dimensional discrete-variable systems \cite{HastrupAndersen}. Also, generation strategies based on ground-state preparation of systems modelled by a GKP Hamiltonian or on dissipation state-engineering have been extensively studied (see Ref. \cite{brady2024advances} and references therein). Recent experiments have succeeded in generating GKP states by a combination of measurements and auxiliary qubits \cite{Fluhmann, CampagneExp, Kudra, deNeeveHome, KonnoGKPgeneration, propagatingGKP}. However, all these methods lie outside the LB approach.

Protocols involving auxiliary qubits and discrete-variable measurements are clearly beyond the scope of sole CV systems, whereas proposals that include ground-state preparation or dissipation have no known decomposition in terms of an LB universal gate set or are not unitary. Implementations involving conditional CV measurements are non-deterministic, and therefore problematic when considering encoded fault-tolerance and the related overhead. 

\begin{figure}[t]
\centering
\colorbox{white}{

\tikzset{every picture/.style={line width=0.75pt}} 

\begin{tikzpicture}[x=0.75pt,y=0.75pt,yscale=-0.57,xscale=0.57]

\draw   (91,9.5) -- (161,9.5) -- (161,49.5) -- (91,49.5) -- cycle ;
\draw   (180.56,9.94) -- (250.56,9.94) -- (250.56,49.94) -- (180.56,49.94) -- cycle ;
\draw   (359.56,10.5) -- (429.56,10.5) -- (429.56,50.5) -- (359.56,50.5) -- cycle ;
\draw    (160.67,30) -- (180.33,30.17) ;
\draw    (339.89,30.78) -- (359.11,30.56) ;
\draw    (80.11,30) -- (90.83,30) ;
\draw    (429.72,30.25) -- (439.11,30.11) ;
\draw  [fill={rgb, 255:red, 0; green, 0; blue, 0 }  ,fill opacity=1 ] (64.61,33.64) .. controls (64.61,33.11) and (65.04,32.68) .. (65.57,32.68) .. controls (66.1,32.68) and (66.53,33.11) .. (66.53,33.64) .. controls (66.53,34.17) and (66.1,34.6) .. (65.57,34.6) .. controls (65.04,34.6) and (64.61,34.17) .. (64.61,33.64) -- cycle ;
\draw  [fill={rgb, 255:red, 0; green, 0; blue, 0 }  ,fill opacity=1 ] (56.78,33.64) .. controls (56.78,33.11) and (57.21,32.68) .. (57.74,32.68) .. controls (58.27,32.68) and (58.7,33.11) .. (58.7,33.64) .. controls (58.7,34.17) and (58.27,34.6) .. (57.74,34.6) .. controls (57.21,34.6) and (56.78,34.17) .. (56.78,33.64) -- cycle ;
\draw  [fill={rgb, 255:red, 0; green, 0; blue, 0 }  ,fill opacity=1 ] (72.41,33.64) .. controls (72.41,33.11) and (72.84,32.68) .. (73.37,32.68) .. controls (73.9,32.68) and (74.33,33.11) .. (74.33,33.64) .. controls (74.33,34.17) and (73.9,34.6) .. (73.37,34.6) .. controls (72.84,34.6) and (72.41,34.17) .. (72.41,33.64) -- cycle ;

\draw    (39.67,30.11) -- (49.83,30.11) ;
\draw    (469.45,30.09) -- (479.33,30.11) ;
\draw  [fill={rgb, 255:red, 0; green, 0; blue, 0 }  ,fill opacity=1 ] (453.63,34.04) .. controls (453.63,33.51) and (454.06,33.08) .. (454.59,33.08) .. controls (455.12,33.08) and (455.55,33.51) .. (455.55,34.04) .. controls (455.55,34.57) and (455.12,35) .. (454.59,35) .. controls (454.06,35) and (453.63,34.57) .. (453.63,34.04) -- cycle ;
\draw  [fill={rgb, 255:red, 0; green, 0; blue, 0 }  ,fill opacity=1 ] (445.8,34.04) .. controls (445.8,33.51) and (446.23,33.08) .. (446.76,33.08) .. controls (447.29,33.08) and (447.72,33.51) .. (447.72,34.04) .. controls (447.72,34.57) and (447.29,35) .. (446.76,35) .. controls (446.23,35) and (445.8,34.57) .. (445.8,34.04) -- cycle ;
\draw  [fill={rgb, 255:red, 0; green, 0; blue, 0 }  ,fill opacity=1 ] (461.43,34.04) .. controls (461.43,33.51) and (461.86,33.08) .. (462.39,33.08) .. controls (462.92,33.08) and (463.35,33.51) .. (463.35,34.04) .. controls (463.35,34.57) and (462.92,35) .. (462.39,35) .. controls (461.86,35) and (461.43,34.57) .. (461.43,34.04) -- cycle ;
\draw   (269.56,10.5) -- (339.56,10.5) -- (339.56,50.5) -- (269.56,50.5) -- cycle ;
\draw    (249.67,30.56) -- (269.33,30.72) ;

\draw (91,15) node [anchor=north west][inner sep=0.75pt]  [font=\large]  {$X( c_i)$};
\draw (182,15) node [anchor=north west][inner sep=0.75pt]  [font=\large]  {$Z( d_i)$};
\draw (363,15) node [anchor=north west][inner sep=0.75pt]  [font=\large]  {$S( r_i)$};
\draw (4.05,18.26) node [anchor=north west][inner sep=0.75pt]  [font=\large]  {$\ket{0}$};
\draw (482.38,16.92) node [anchor=north west][inner sep=0.75pt]  [font=\large]  {$\ket{\psi _{g}}$};
\draw (271,15) node [anchor=north west][inner sep=0.75pt]  [font=\large]  {$K\!(k_i )$};

\end{tikzpicture}
}
\caption{One block of the circuit of universal CV gates used for generating GKP states. Here $X(c_i)$ is a displacement in position, $Z(d_i)$ is a displacement in momentum, $K(k_i)$ is the Kerr gate and $S(r_i)$ is squeezing. The input is the vacuum state $\ket{0}$ and the output is the generated state $\ket{\psi_g}$.} \label{fig:circuit}
\end{figure}
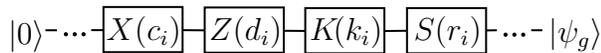   

Therefore, we employ a numerical approach to generate GKP states in a deterministic, unitary and fully CV manner. We use a parameterized quantum circuit inspired by the approach introduced by Arrazola \textit{et al.} in Ref.~\cite{arrazola2019machine}. Specifically, we consider a circuit made up of the repeating blocks shown in Fig.~\ref{fig:circuit}, 
each one composed in turn of gates included in the universal set in Eqs.~(\ref{Xgate}-\ref{Kgate}) or straightforwardly obtainable from them \footnote{In particular, the momentum displacement gate is defined as $Z(d)=\exp\left[id\hat{q}\right]$, and can be obtained as a combination of the $X(c)$ and $R(\theta)$ gates. In addition, the Kerr gates $K(k)$ and $K_{LB}(k)$ are related via a rotation: $K(k)\propto K_{\rm{LB}}(k/4)R(-k)$.}. The gate parameters are given by the vectors $\boldsymbol{c}, \boldsymbol{d}, \boldsymbol{k}, \boldsymbol{r}$, each of dimension equal to the total number of blocks, where the parameters for the $i$-th block are given by the $i$-th element of each of these vectors. We take the vacuum as input, whereas we set the target state  $\ket{\Psi_t}$ as the square GKP $\ket{\Tilde{0}}$ in Eq.~(\ref{eq:approxgkp}). We use the Strawberry Fields Python library \cite{SF1killoran2019,SF2bromley2020} to simulate this circuit using a truncated Fock representation. We optimize the gate parameters in order to maximize the fidelity, $F=|\bra{\Psi_g}\ket{\Psi_t}|^2$, between the generated and the target state.

\begin{figure}[t]
\center{\includegraphics[width=0.8\textwidth]{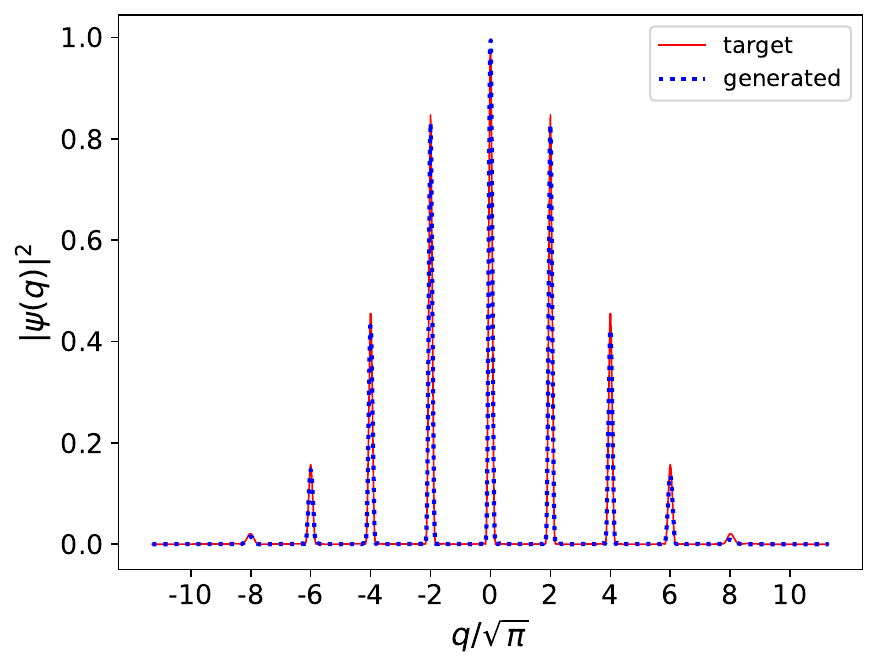}}
\caption{Probability distributions for a target $\ket{\Tilde{0}}$ state, with $\Delta=0.12$, and the state generated numerically using a Fock basis cutoff of 120.}
\label{fig:ProbDists}
\end{figure}

An example of the generated states compared to the target can be seen in Fig.~\ref{fig:ProbDists}, which shows the position probability distributions for both, for $\Delta=0.12$. For this case, the fidelity between the target and generated states is 0.987. More in general, we generate states with inverse width $\Delta$ between 0.4 and 0.08, obtaining infidelities, $1-F$, below 0.03 as shown in Fig.~\ref{fig:PerrorVsDelta} (dashed line, right axis). Details of our simulation can be found in Appendix A and the code is available at Ref.~\cite{GKPfromvac}. 

Now that we have seen that the GKP qubits $\ket{\Tilde{0}}$ can be prepared approximately within an LB universal model, encoded universality can be proven by showing how to implement universal circuits on the embedded qubits. A conventional choice \cite{Nielsen&Chuang} for the latter circuits comprises \textit{(i)} the preparation and measurement in the qubit computational basis; \textit{(ii)} the ability to implement arbitrary encoded Clifford circuits as sequences of Hadamard $\bar H$, phase $\bar S$, and control-$\bar Z$ gates; \textit{(iii)} the ability to implement an encoded non-Clifford gate such as the $\bar T$ gate. Therefore, encoded universality follows from three considerations. First, encoded basis measurements correspond to homodyne measurements with binned outcomes, which are already within our minimally-extended LB universal model. Second, encoded Clifford gates can be obtained immediately from the chosen LB universal set, since the control-$\bar Z$ gate coincides with $CZ$, the Hadamard $\bar H$ gate corresponds to $R(\pi)$, and the phase gate $\bar{S}(s)$ corresponds to $R(\theta)S(r)R(\theta-\pi/2)$, where \cite{kalajdzievski_quesada_2021}:
\be
s=2\sinh{r}\,; \;\; \cos{\theta}=\frac{1}{\sqrt{1+e^{2r}}}\,; \;\; \sin{\theta}=\frac{e^r}{\sqrt{1+e^{2r}}}\,.
\ee
Third, the $\bar T$ gate can be approximately implemented using encoded Clifford gates along with a supply of encoded $\ket{\Tilde{\scalebox{0.9}{$H$}}}$ states \cite{koashi_GaussUniv}, where
\begin{equation}
    \ket{\Tilde{\scalebox{0.9}{$H$}}}=\frac{1}{\sqrt{2}}\left(e^{-i\frac{\pi}{8}}\ket{\Tilde{0}}+e^{i\frac{\pi}{8}}\ket{\Tilde{1}}\right)\, .
\end{equation}
We can generate these $\ket{\Tilde{\scalebox{0.9}{$H$}}}$ states using the same method as for the $\ket{\Tilde{0}}$ states, obtaining similar fidelities, as shown in Appendix A. Alternatively, notice that one could implement a $\bar T$ gate simply by equipping the reference LB universal set in Eqs.~(\ref{Xgate}-\ref{Kgate}) with an additional cubic-phase gate \cite{gkp}.

\textit{Fault-Tolerance---} We now address whether the LB notion of computational universality is compatible with fault tolerance. Specifically, we argue that a fault-tolerant quantum memory can indeed be realized using GKP-encoded qubits prepared via the LB universal gate set defined in Eqs.~(\ref{Xgate}-\ref{Kgate}). Our primary objective is to show that the imperfections inevitably introduced by the LB framework during state preparation do not, in themselves, preclude fault tolerance. To isolate the effect of encoding imperfections, we assume that all operations from the LB universal set used in the error correction procedure are implemented ideally, with no additional sources of error.

Specifically, our argument builds on the results of Ref.~\cite{NohChambBrand2022}, where it was shown that infinite-energy GKP states can serve as qubits in a surface code architecture to realize a quantum memory with an error rate below the fault-tolerance threshold. Therefore, to establish that LB universal sets enable fault-tolerant quantum error correction, it suffices to demonstrate that such a set can generate a supply of GKP states which, despite inherent imperfections and finite energy, are of sufficient quality to meet the conditions for fault-tolerance.

To this aim, we calculate the probability of finding uncorrectable errors upon measuring the quadratures of the GKP states generated with the procedure presented above. Errors are deemed uncorrectable if they do not satisfy the Glancy-Knill condition \cite{glancyknill}, which provides an upper bound on the size of shift errors that can be corrected by repeated rounds of Steane-type GKP error correction. We then compare this probability of incorrectly identifying the logical states with the threshold of some discrete-variable error-correction code. To the best of our knowledge, this is the first time that an error analysis has been performed for approximate GKP states that can be prepared with a physically realizable procedure prescribed within the LB framework.

\begin{figure}[t]
\centering
\scalebox{1.2}{
\,
\Qcircuit @C=2em @R=1.7em {
 \lstick{ \ket{\Tilde{\mu}}}             & \targ & \qw                & \gate{Z(-p\,\textrm{mod}\sqrt{\pi})}   & \qw \\
 \lstick{\ket{\Tilde{0}}}  & \ctrl{-1}        & \measureD{\hat{p}} & \control \dstick{p} \cw  \cwx 
}
}
\caption{Quantum circuit for correcting momentum errors in an approximate GKP state, $\ket{\Tilde{\mu}}$, using an auxiliary GKP qubit, $\ket{\Tilde{0}}$ \cite{gkp,glancyknill}. An inverse-SUM gate, $\exp[i\hat{q}_1\hat{p}_2]$, acts on the inputs. Then the second mode is measured in the momentum basis and the measurement outcome, $p$, is fed to a displacement gate, $Z(-p\,\textrm{mod}\sqrt{\pi})$. Notice that all the elements of the circuit are comprised in our LB universal model.}
\label{fig:ECcircuit}
 \end{figure}
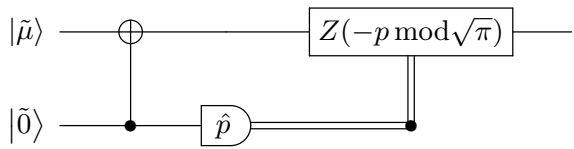

Consider the protocol for correcting a momentum error in an approximate GKP state, $ \ket{\Tilde{\mu}}$, as shown in Fig.~\ref{fig:ECcircuit}. To correct errors in position we use this circuit again after a Fourier transform. The Glancy-Knill condition \cite{glancyknill} states that repeated use of this quantum error correction procedure succeeds when the magnitude of each displacement error measured is less than $\sqrt{\pi}/6$. In this context, failure means that either a logical error is introduced or the quality of the data register is decreased by the process of syndrome extraction with imperfect auxiliary states and subsequent correction. The probability that, upon measurement, a general state $\ket{\psi}$ gives outcomes larger than $\sqrt{\pi}/6$ is given by
\be
P(\textrm{error})=1-\int_{-\sqrt{\pi}/6}^{\sqrt{\pi}/6}{du}\int_{-\sqrt{\pi}/6}^{\sqrt{\pi}/6}{dv \abs{\innpro{u,v}{\psi}}^2}\,,
\label{eq:Perror}
\ee
where $\ket{u,v}$ is the state shifted by $u$ in position and $v$ in momentum from the ideal $\ket{\overline{0}}$, i.e.,
\be
\ket{u,v}=\pi^{-1/4}e^{-iu\hat{p}}e^{-iv\hat{q}}\ket{\overline{0}}\,.
\ee
For fault-tolerance we require $P(\textrm{error})<P_{th}$, where $P_{th}$ is the threshold value for a given discrete-variable error correcting code operating at the logical level, for a given noise model. 

In Ref.~\cite{NohChambBrand2022}, Noh et al. showed that when GKP error correction is concatenated with the surface code \cite{surfacecode} to correct errors at the qubit level, the dominant noise source is the finite squeezing of the auxiliary GKP states used for physical error correction. They made use of the twirling approximation \cite{menicucci}, in which approximate GKP states are modelled as ideal GKP states with the Dirac delta functions replaced by Gaussians with constant non-zero variance. Unlike the approximate GKP states given in Eq.~(\ref{eq:approxgkp}), this approximation does not include a surrounding Gaussian envelope and hence the states are unphysical (infinite energy) as their wavefunctions extend to infinity in position and momentum, but this greatly simplifies the error analysis. In particular, this approximation allows for the estimation of the logical error probability of the \emph{overall} concatenated error correction procedure using Monte Carlo methods. The authors found a threshold squeezing of 9.9dB for fault-tolerant quantum error correction with twirling-approximated GKP states. We note that this threshold is in relatively close agreement with the 10.5dB threshold found in Ref.~\cite{bourassa2021} for a measurement-based fault-tolerant quantum memory \footnote{This threshold of 10.5dB was considered in Ref.~\cite{marek}, where it was converted to a threshold expressed in terms of a measure of the non-Gaussianity of GKP states which they introduce, called ``GKP squeezing''.}.

Here we take the 9.9dB squeezing threshold and use it to identify a threshold $P_{th}$ in terms of error probability. Following the reasoning of Ref.~\cite{NohChambBrand2022}, we assume that, when using imperfect generated GKP states as the qubits in a surface code, the dominant sources of noise are the finite squeezing of the GKP peaks and the imperfect fidelity between the target and generated GKP states. Both of these factors are taken into account by the probability of error in Eq.~(\ref{eq:Perror}). In other words, we assume that the quantity in Eq.~(\ref{eq:Perror}) is sufficient to characterize the effective qubit channel induced on the input to the next layer of encoding, which might for example be the rotated surface code considered in \cite{NohChambBrand2022} or the measurement-based code in~\cite{bourassa2021}. This is justified by the observation that decoding of concatenated GKP codes is typically performed in two steps: first GKP-level stabilizers are measured, essentially converting CV noise to qubit-level noise, then qubit-level stabilizers are measured and a higher level correction is computed. The second step succeeds if the strength of the effective qubit-level noise is low enough after the first step, which is the essence of the Glancy-Knill criterion. Strictly speaking, $P(\mathrm{error})$ alone does not characterize the most general qubit channel, but since the qubit-level noise is local by construction and symmetry between bit-flips and phase-flips is approximately ensured by the large fidelity with the target state (which is symmetric), we can assume that the true effective qubit-channel will not be significantly more detrimental to the operation of the higher level error correction than a stochastic bit-phase flip channel of equal intensity.

In order to obtain the value of the fault-tolerant threshold $P_{th}$ for the GKP surface code with this noise model, we need to determine $P(\textrm{error})$ for the twirling-approximated GKP states with squeezing 9.9dB (namely, $\Delta=0.32$). We make use of the fact that twirling-approximated GKP states are periodic, with periodicity $2\sqrt{\pi}$. Hence the ratio $P(\textrm{error}):P(\textrm{no error})$ for their full infinitely extended wavefunction is equal to that ratio for just one of the repeating cells of that wavefunction, $q\in\left(-\sqrt{\pi},\sqrt{\pi}\,\right]$. This allows us to calculate $P(\textrm{error})$ for different values of $\Delta$. The results are plotted in Fig.~\ref{fig:PerrorVsDelta} (dot-dashed line). In particular, we see that the value for $\Delta=0.32$ is $P_{th}=0.347$, which is also shown in Fig.~\ref{fig:PerrorVsDelta} (horizontal line). Notice that the error probability for the target states in Eq.~(\ref{eq:approxgkp}) is also plotted (dotted line), and it is almost indistinguishable from the one corresponding to the twirling-approximated states. This confirms the validity of the twirling approximation in the range of $\Delta$ values shown \footnote{A slight divergence can be seen between the error probability for target and twirling states for $\Delta$>0.35, or equivalently, for squeezing below 9.12dB. This is due to the fact that, in the coherent-envelope case of the target state, the peaks far from the origin are slightly displaced. Even though the envelope operator does not deform phase space, it affects the value of the Wigner function more at points which are further from the origin, which causes an apparent displacement of the peaks. The twirled states are not affected by this.}, corroborating the results found in Ref.~\cite{hillmann}. Therefore we conclude that a fault-tolerant memory can be achieved if it is possible to generate GKP states with $P(\textrm{error})$ significantly smaller than $P_{th}=0.347$.

The identified threshold can now be used to assess the quality of the states generated within the LB model, as well as any state in general. The error probability for the generated states is plotted in Fig.~\ref{fig:PerrorVsDelta} as well (solid line). We see that in the region below $\Delta=0.3$ (squeezing above 10.5dB), the error probability for the generated states is below the fault-tolerance threshold, $P_{th}$. The lowest error probability obtained is 0.0174, which occurs at $\Delta=0.12$, significantly lower than the threshold \footnote{Notice that for very high squeezing ($\Delta < 0.12$) the error probability of our generated states starts to increase again, because the fidelity between the generated and target GKP states starts to decrease in this parameter regime. This is due to the fact that these highly squeezed states require a large cutoff and so there are more Fock components that need to be optimized, which increases the run-time. We have chosen the number of blocks and number of trials to allow the optimization to run in an acceptable time frame, but these are too low to achieve the same high fidelities obtained for lower values of squeezing (higher $\Delta$).}.  Therefore, we can conclude that the states generated using the LB framework can be of sufficient quality for a fault-tolerant quantum memory. Note that here we only consider fault-tolerance in the error correction of the prepared GKP states. In particular, we do not analyse how gates could lead to error propagation. Hence, as said, our result does not show fault-tolerant quantum computation, but instead is a demonstration of a fault-tolerant quantum memory.

\begin{figure}[t]
\center{\includegraphics[width=\textwidth]{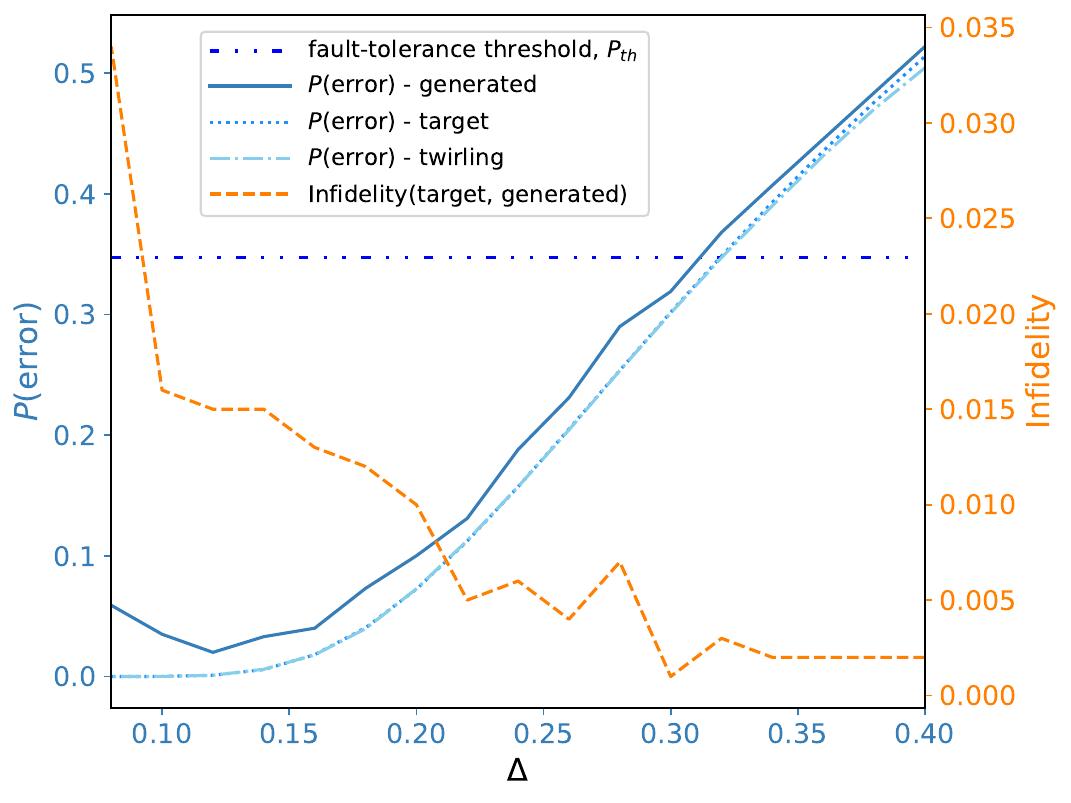}}
\caption{Error probability, calculated using Eq.~(\ref{eq:Perror}), for target GKP $\ket{\Tilde{0}}$ and generated states with different values of $\Delta$, the inverse width. We show the error probability for the twirling states used in Ref.~\cite{NohChambBrand2022}, as well as the fault-tolerance threshold. The infidelity between the target and generated GKP states is also plotted. }
\label{fig:PerrorVsDelta}
\end{figure} 

To assess the usefulness of our generated states for quantum computation, even in the case of assuming perfect gates, we would need a way to take into consideration the fact that when we apply perfect gates to our imperfect (finite squeezing and non-unit fidelity) states, these gates do not act perfectly \cite{Konno2021NonCliffor, Hastrup2021Unsuitability}. For example, after applying a perfect $X$ gate to our generated GKP $\ket{\Tilde{0}}$ with $\Delta=0.12$ from Fig.~\ref{fig:ProbDists}, the error probability for the resulting GKP $\ket{\Tilde{1}}$ is 0.0313, compared to 0.0197 for the original $\ket{\Tilde{0}}$ state. Although both of these values are well below the fault-tolerance threshold, further analysis would be required to determine whether GKP states generated in this way can facilitate full fault-tolerant quantum computation.

\textit{Conclusions---} We have investigated a computational framework regarded as universal in the CV setting, according to the seminal notion introduced by Lloyd and Braunstein \cite{BraunLloyd}, and recently experimentally realized \cite{eriksson2024universal}. Within this framework, we have demonstrated the ability to generate encoded GKP qubits suitable for performing universal encoded computation and, crucially, that these qubits can support a fault-tolerant quantum memory. Specifically, under the concatenated GKP-surface code and noise model considered in Ref.~\cite{NohChambBrand2022}, and assuming the only source of noise stems from state preparation imperfections, we derived a threshold for the maximum tolerable error probability. We found that targeting GKP qubits with squeezing exceeding 10.5 dB, and maintaining high fidelity to this target, the generated states yield error rates below this threshold.

These results show that the long-standing Lloyd–Braunstein framework is not only computationally universal in the original sense of Ref.~\cite{BraunLloyd} but it is also capable of supporting fault-tolerant quantum memories, a key requirement for scalable quantum computation. This marks a significant theoretical advancement in establishing the practical viability of the Lloyd–Braunstein approach to CV quantum computing. To fully demonstrate its compatibility with fault-tolerant quantum computation, future work should analyze how imperfections in the GKP qubits propagate through logical gate implementations within this framework.

{\it Acknowledgements} A.F.\ and G.F.\ acknowledge funding from the European Union’s Horizon Europe Framework Programme (EIC Pathfinder Challenge project Veriqub) under Grant Agreement No.\ 101114899.
G.F.\ acknowledges financial support from the Swedish Research Council (Vetenskapsrådet) through the project grant DAIQUIRI and from the Knut and Alice Wallenberg Foundation through the Wallenberg Center for Quantum Technology (WACQT). S.B. acknowledges support from the UK EPSRC through Grant No. 2442912.

\section{Appendix A: Simulation details}
\label{sec:appendix}

As said in the main text, we employ a numerical approach to generate GKP states inspired by Ref.~\cite{arrazola2019machine}. Specifically, we consider a circuit made up of the repeating blocks shown in Fig.~\ref{fig:circuit}. We use the Strawberry Fields Python library \cite{SF1killoran2019,SF2bromley2020} to simulate this circuit and optimize the gate parameters in order to maximize the fidelity between the generated and the target state. Notice that, although the algorithm optimizes the fidelity, this metric alone is not always a sufficiently sensitive indicator of performance, especially for CVs \cite{bina2014drawbacks, mandarino2014use}. Accordingly, the quality of the generated quantum state with respect to fault tolerance is evaluated using the more comprehensive methods described in the main text. We run the optimization several times, sampling different starting parameters at random each time and take the best output state. Before optimization can begin, several hyperparameters must be set, including the number of blocks in the circuit, the number of trials of each simulation, and the cutoff (the dimension at which the Fock basis is truncated).  
 
\begin{figure}[t]
\center{\includegraphics[width=\textwidth]{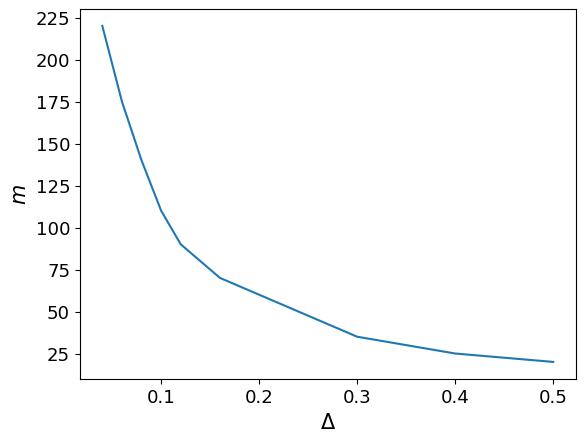}}
\caption{Cutoff dimension at which to truncate the Fock basis for various values of the parameter $\Delta$ that characterizes the target GKP states.}
\label{fig:BestCutoffVsDelta}
\end{figure}

The choice of cutoff depends on the value of $\Delta$ of the target state. To set the best (minimal) cutoff, we calculated the fidelity between a GKP state with cutoff $n$ and that same GKP state with cutoff $n+1$. For low values of $n$, this fidelity oscillates rapidly, but by increasing $n$ we can find a cutoff, $m$, beyond which the difference between the states with cutoff $m$ and $m+1$ is negligible ($F>0.999$). This reveals the dimension at which we can truncate the Fock space whilst allowing almost all of the information about the state to be preserved. The best cutoff $m$ found for various values of $\Delta$ is shown in Fig.~\ref{fig:BestCutoffVsDelta}.

We target GKP $\ket{\Tilde{0}}$ states with inverse width $\Delta$ between 0.4 and 0.08. For the highest values of $\Delta$, which corresponds to the lowest squeezing, we use a circuit with 10-20 blocks and cutoff of 30. We need to increase the cutoff as we decrease $\Delta$ (see Fig.~\ref{fig:BestCutoffVsDelta}) since this means we are targeting more complex GKP states. Such states are more challenging to generate and so we also increase the number of blocks in our circuit, using 60 blocks for $\Delta=0.08$, and a cutoff of 150. To check that the optima we find are numerically stable, and not artifacts of the cutoff, we run the optimal circuit with a higher cutoff than that used in optimization. We see that the state remains close to normalized, with $\bra{\Psi_g}\ket{\Psi_g}$>0.996 for the higher cutoff.

\begin{figure}[h]
\center{\includegraphics[width=\textwidth]{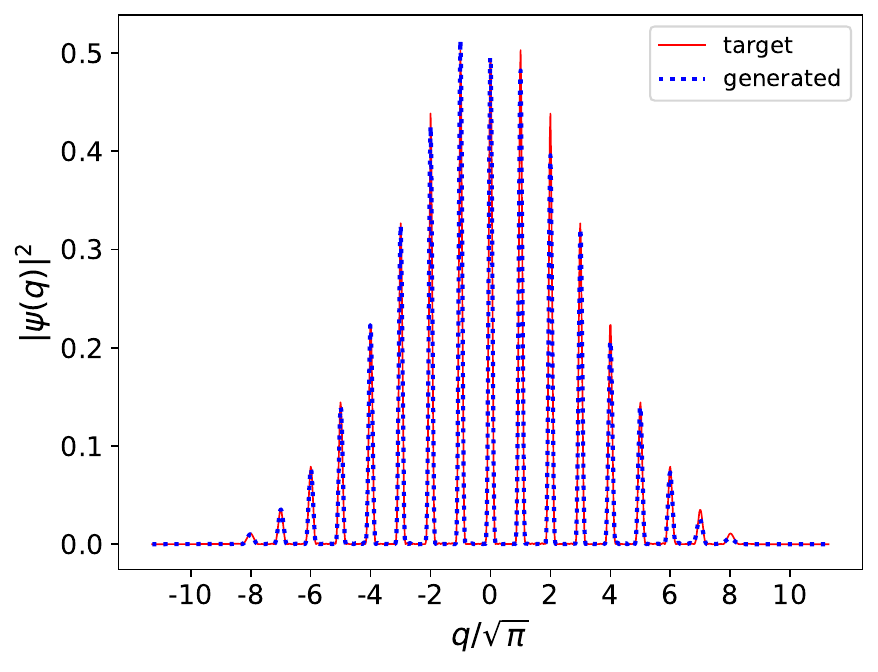}}
\caption{Probability distributions for the numerically generated and the target GKP $\ket{\Tilde{\scalebox{0.9}{$H$}}}$ state with the same parameters as for the $\ket{\Tilde{0}}$ state, i.e. $\Delta=0.12$ and a cutoff of 120. The fidelity between the two states is $F=0.991$.}
\label{fig:Hstate}
\end{figure}

With the same technique, it is possible to generate deterministically the GKP $\ket{\Tilde{\scalebox{0.9}{$H$}}}$ from the vacuum with the elementary gates belonging to the LB model in Fig.~\ref{fig:circuit}. As an example, in Fig.~\ref{fig:Hstate} we present the resulting probability distribution in the position representation for a target GKP $\ket{\Tilde{\scalebox{0.9}{$H$}}}$ state with $\Delta=0.12$, yielding a fidelity $F = 0.991$. The optimized circuit parameters leading to this state are available at Ref.~\cite{GKPfromvac}.

\providecommand{\noopsort}[1]{}\providecommand{\singleletter}[1]{#1}%

\end{document}